# Respiratory Anomaly Detection using Reflected Infrared Light-wave Signals

Md Zobaer Islam*, Brenden Martin†, Carly Gotcher†, Tyler Martinez†, John F. O'Hara†, and Sabit Ekin‡

*Department of Radiology, University of North Carolina, Chapel Hill, NC, USA
†School of Electrical and Computer Engineering, Oklahoma State University, Stillwater, OK, USA
‡Department of Engineering Technology and Industrial Distribution, Texas A&M University, College Station, Texas, USA
E-mail: zobaer_islam@med.unc.edu, {brenden.martin, carly.gotcher, tyler.martinez, oharaj}@okstate.edu, sabitekin@tamu.edu

*Abstract*— In this study, we present a non-contact respiratory anomaly detection method using incoherent light-wave signals reflected from the chest of a mechanical robot that can breathe like human beings. In comparison to existing radar and camera-based sensing systems for vitals monitoring, this technology uses only a low-cost ubiquitous infrared light source and sensor. This light-wave sensing system recognizes different breathing anomalies from the variations of light intensity reflected from the chest of the robot within a 0.5m-1.5m range with an average classification accuracy of up to 96.6% using machine learning.

## Introduction

Human respiratory rate and its pattern convey useful information about the physical and mental states of the subject. Contact-based respiration monitoring is not effective enough in capturing the true essence of one's breathing, because people may undergo a shift in breathing pattern when they know that they are being monitored. Moreover, the subject might be too young or too ill to apply contact-based approaches for a longer duration. Existing automated technologies for non-contact breathing monitoring are typically based on sensing electromagnetic signals (radio frequency, WiFi etc.) or extracting breathing information from videos captured by red-green-blue (RGB) or thermal infrared (IR) cameras. We have developed a non-contact respiration monitoring and anomaly detection system that is based on sensing incoherent infrared light reflected from the subject's chest. This technology is superior to the existing technologies because of the safe, ubiquitous and discreet nature of infrared light and the absence of privacy issues [1].

## System Design, Implementation and Results

A mechanical robot was developed to mimic normal and abnormal human breathing patterns precisely for this study. The system comprises light-wave sensing (LWS) hardware, data collection and storage units, signal processing algorithms, and classification algorithms (see Fig. 1). IR light-emitting diode sources illuminate the subject's chest, and the reflected light intensity, varying with chest movement, is converted into a time-series electrical signal. 60s long data collected at 0.5m, 1m and 1.5m distances were processed, features extracted, and classified using Decision Tree (DT) and Random Forest (RF) machine learning models. The classes and their characteristics are detailed in Table 1, with breathing rates in breaths per minute (BPM) and breathing depth as a percentage of maximum rib cage movement [2]. Class 7 identifies erroneous data lacking breathing waveforms for rejection or recollection.

Test accuracies, with 10-fold cross-validation (CV), are presented in Table 2, indicating >94% accuracy within 1m test distance with both models. Accuracy decreased beyond 1m due to higher path-loss and lower signal-to-noise ratio in the received signal. In such case, RF, an ensemble ML model, outperformed a single decision tree in classifying the respiration data.

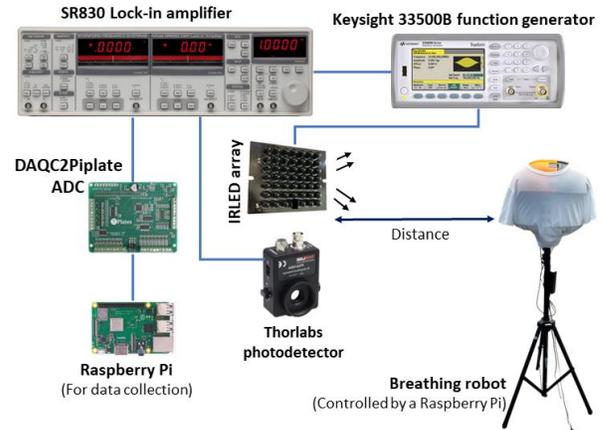

Fig. 1: Experimental setup

TABLE 1: Characteristics of breathing classes

|   | Classes | Breathing rate (BPM) | Breathing Depth (%) |
|---|---------|----------------------|---------------------|
| 0 | Eupnea  | 12-20 | 30-58 |
| 1 | Apnea   | 0 | 0 |
| 2 | Tachypnea | 21-50 | 30-58 |
| 3 | Bradypnea | 1-11 | 30-58 |
| 4 | Hyperpnea | 12-20 | 59-100 |
| 5 | Hypopnea | 12-20 | 1-29 |
| 6 | Kussmaul's | 21-50 | 59-100 |
| 7 | Faulty data | Any | Any |

TABLE 2. Classification accuracies

|    | Average classification accuracies (CV-based) | | |
|----|------|------|------|
|    | 0.5m | 1m   | 1.5m |
| DT | 96.6% | 94.1% | 71.9% |
| RF | 96.3% | 94.1% | 75.3% |

## Acknowledgment

This work was supported by the National Science Foundation under Grants 2008556 and 2050062.